\documentclass{aa}
\usepackage{graphics}

\begin{document}

   \thesaurus{04(04.03.1; 04.19.1; 11.13.1; 11.16.1, 11.19.5; 13.09.6)
              } 

   \title{The DENIS Point Source Catalogue towards the Magellanic Clouds
   \thanks {Based on observations collected at the European Southern 
            Observatory}}

   \subtitle{}

   \author{M-R. Cioni\inst{1}
           \and
	   C. Loup\inst{2}
	   \and
	   H.J. Habing\inst{1}
           \and
           P. Fouqu\'e\inst{6,3}  
	   \and
	   E. Bertin\inst{2}
	   \and
	   E. Deul\inst{1}
	   \and
	   D. Egret\inst{14}
	   \and
	   C. Alard\inst{15}
	   \and
	   B. de Batz\inst{3}
	   \and
	   J. Borsenberger\inst{2}
	   \and 
           M. Dennefeld\inst{2}
	   \and
           N. Epchtein\inst{4}
	   \and  
           T. Forveille\inst{5} 
	   \and
           F. Garz\'on\inst{7}
           \and
	   J. Hron\inst{8} 
           \and
           S. Kimeswenger\inst{9} 
           \and
	   F. Lacombe\inst{3}
           \and
	   T. Le Bertre\inst{15}  
           \and
 	   G. A. Mamon\inst{2,3} 
           \and
	   A. Omont\inst{2} 
           \and
	   G. Paturel\inst{10} 
           \and
	   P. Persi\inst{11} 
           \and
	   A. Robin\inst{12}
           \and
	   D. Rouan\inst{3} 
           \and
	   G. Simon\inst{15}
           \and
	   D. Tiph\`ene\inst{3}
           \and
	   I. Vauglin\inst{10}
	   \and 
	   S. Wagner\inst{13}          
	}

   \offprints{M-R. Cioni}

   \institute{Leiden Observatory, Postbus 9513,
              NL--2300 RA Leiden, The Netherlands\\
         \and
             Institut d'Astrophysique de Paris (CNRS UPR 341), 
	     98 bis Bd. Arago, F--75014 Paris, France\\
	 \and
	     Observatoire de Paris, 
	     5 place J. Janssen, F--92195 Meudon Cedex, France\\
	 \and
	     Observatoire de la C\^ote d'Azur, D\'epartement Fresnel, 
	     F--06304 Nice Cedex 04, France\\
	 \and
	     Observatoire de Grenoble, 414 rue de la Piscine, 
             Domaine Universitaire de Saint Martin d'H\`eres, 
             F-38041 Grenoble, France\\
	 \and
	     European Southern Observatory, 
	     Casilla 19001, Santiago 19, Chile\\
	 \and
	     Instituto de Astrofisica de Canarias
	     E--38200 La Laguna, Tenerife, Spain\\
	 \and
	     Institut f\"{u}r Astronomie der Universit\"{a}t Wien,
             Turkenschanzstrasse 17, A--1180 Wien, Austria\\
	 \and
	     Institut f\"{u}r Astrophysik, Innsbruck University
	     A--6020 Innsbruck, Austria\\
	 \and
	     CRAL--Observatoire de Lyon
	     F--69561 Saint--Genis Laval Cedex, France\\
	 \and
	     Istituto di Astrofisica Spaziale,
	     Area di Ricerca Roma--Tor--Vergata, I--00044 Italy\\ 
	 \and
	     Observatoire de Besan\c{c}on
	     BP 1615, F-25010 Besan\c{c}on Cedex, France\\
	 \and
	     Landessternwarte Heidelberg,
	     K\"onigstuhl, D--69117 Heidelberg, Germany\\
         \and
	     CDS, Observatoire Astronomique de Strasbourg (CNRS UMR 7550),
	     11 rue de l'Universit\'e, F--67000 Strasbourg, France\\    
	 \and
	     Observatoire de Paris, 65'' Avenue de l'Observatoire, 
             F--75014 Paris, France\\
		}

   \date{Received 6 December 1999 / Accepted 10 March 2000}

   \titlerunning{The DENIS PSC towards the Magellanic Clouds}
   \authorrunning{M-R. Cioni et al.}

   \maketitle

   \begin{abstract} We have compiled  the near infrared Point
Source Catalogue  (PSC) towards the Magellanic  Clouds (MCs) extracted
from  the data  obtained with  the Deep  Near Infrared  Survey  of the
Southern Sky  -- DENIS (Epchtein et al.   \cite{ept1}).  The catalogue
covers  an area  of of  $19.87\times  16$ square  degrees centered  on
$(\alpha,\delta)=(5^h27^m20^s$,$-69\degr00\arcmin00\arcsec)$  for  the
Large  Magellanic  Cloud  (LMC)  and $14.7\times  10$  square  degrees
centered                                                             on
$(\alpha,\delta)=(1^h02^m40^s$,$-73\degr00\arcmin00\arcsec)$  for  the
Small  Magellanic  Cloud  (SMC) at the epoch J2000.   
It contains  about $1\,300\,000$  sources
towards the LMC  and $300\,000$ sources towards the  SMC each detected
in at least 2 of the  3 photometric bands involved in the survey ($I$,
$J$, $K_s$).  $70\%$  of the detected sources are  true members of the
Magellanic  Clouds, respectively  and  consist mainly  of red  giants,
asymptotic giant branch stars and super-giants.  The observations have
all  been  made with  the  same  instrument  and 
the data have been calibrated and reduced uniformly.  
The catalogue  provides  a homogeneous set of photometric data.
	
   \end{abstract}

\section{Introduction}
The  DENIS  project aims  to  survey  the  entire southern  hemisphere
simultaneously in  three photometric  bands, $I$ (Gunn--i,  $0.8\, \mu
m$), $J$  ($1.25\, \mu m$) and  $K_s$ ($2.15\, \mu m$)  with a spatial
resolution of  $1\arcsec$ in $I$ and  $3\arcsec$ in the  $J$ and $K_s$
bands,  and  limiting magnitudes  of  $I=18$,  $J=16$, $K_s=14$.   See
Epchtein et al. (\cite{epal}) for  the first general release of DENIS
data. Here  we  present a  catalogue  of  DENIS  Point Sources  towards  the
 Magellanic Clouds,  requiring that objects  are detected in  at least
 two of the three photometric bands. At the distance of the Magellanic
 Clouds,  $(m-M)=18.45\pm0.1$ for the  LMC and  $(m-M)=19.0\pm0.1$ for
 the  SMC  according   to  Westerlund  (\cite{west}),  our  catalogue
 contains: (1)  all Asymptotic Giant Branch stars  (AGB), except those
 with shells optically thick at $2\mu m$ and the faintest stars at the
 very beginning of  the Early AGB branch (E-AGB),  (2) upper Red Giant
 Branch  stars  (RGB),  (3)  most  of the  super-giants  except  those
 brighter than $I=10.5$, $J=8.0$,  $K_s=6.5$ because they saturate the
 detectors,  
 (4) relatively bright post-AGB  stars. The catalogue will thus be
 a  major  tool for  statistical  studies  of  the post-main  sequence
  stellar  populations of  the Magellanic Clouds.  Dwarfs and
 giants are the main galactic  sources seen in front of the Magellanic
 Clouds   (Ruphy   et   al.   \cite{ruph}).   Compared   to   earlier
 spectroscopic and  photometric surveys  of the Magellanic  Clouds for
 red giants  and super-giants, and for  stars on the  AGB, probably we
 find  a few  hundreds  times more  sources, for several reasons:
(1)  previous
 surveys  were   not  sensitive  enough   (Westerlund  \cite{west0},
 \cite{west1}; Sanduleak \&  Philip \cite{saph}; Westerlund et al.
 \cite{weal1},  \cite{weal2};  Rebeirot  et al.  \cite{reb0}), 
 (2) they were spatially  limited (see e.g.  Blanco, McCarthy, and
 Blanco \cite{bmc}; Blanco \& McCarthy \cite{blamc}), (3) they
 were  restricted  to  a  peculiar  type  of  objects  (e.g.   Hughes
 \cite{hug}  in  his  search  for  Miras variables,  Rebeirot  et
 al. \cite{reb} in their search for carbon stars). About $1/4$ of the
 sources  discovered  in these  surveys  were  later  observed in  the
 $JHK(L)$   infrared   photometric  bands   (e.g.    Hughes  \&   Wood
 \cite{hugwo},  Costa  \&  Frogel  \cite{cofrog}).   DENIS  provides
 simultaneous $IJK_s$  observations of the entire Clouds,  with a good
 sensitivity,  and  connecting  for  the first  time  the  traditional
 optical and infrared wavelengths domains by simultaneous observations.

Sect. $2$ describes the instrument characteristics and the observing
technique.  Sect. $3$ describes  the data reduction procedure in the
two ``data analysis centers``  with particular attention to: flat and
bias   subtraction,  point  spread   function,  and   astrometric  and
photometric  calibration.  Sect.  $4$ discusses  the quality  of the
data  with  regard  to  the  selection criteria  applied  and  to  the
completeness  reached.   Sect.  $4.2$  discusses in  particular  the
foreground sources belonging  to our  Galaxy.  Finally,  Sect. $5$
describes  the  content  of   the  catalogue  and  Sect.  $6$  gives
conclusive remarks. The catalogue is available through the Strasbourg 
Astronomical Data Center (CDS); it carries the number II/228.

\section{Observations}
The DENIS  instrument is mounted at  the Cassegrain focus  of the 1--m
ESO  telescope (La  Silla  -  Chile).  It  contains  three cameras:  a
Tektronix  CCD with  $1024\times1024$ pixels  and two  NICMOS infrared
detectors  with $256\times256$ pixels.   The array  of the  camera has
four  quadrants  to  reduce  the  read-out time,  and  each  quadrant,
especially in  the $I$ band, presents  different image characteristics
and must be treated separately.  The pixel sizes are $1\arcsec$ in $I$
and $3\arcsec$ in $J$  and $K_s$, respectively.  The total integration
time  is $9$  secs  for each  image.   The sampling  of  the image  is
$1\arcsec$  in all three  wave bands.   The $J$  and $K_s$  images are
dithered  to  a $1\arcsec$  pseudo--resolution,  using a  microscanning
mirror.  They consist of a set  of $9$ frames each obtained in $1$ sec
integration time, shifted  by $\pm 1/3$ pixel in  right ascension (RA)
plus $\pm 7/3$ pixel in declination (DEC).

The DENIS strategy  is to divide the sky  into three declination zones
and scan  each in  strips of $30\degr$  in DEC and  $12\arcmin$ in
RA. The overlap in RA  between consecutive strips is $2\arcmin$.  Each
strip  consists of  180 images  of $12\arcmin\times12\arcmin$  with an
overlap  of  $2\arcmin$ between  each  image.   The  observation of  a
photometric standard star consists of $8$ sub-images shifted according
to a circular pattern in order  to have the star always at a different
position  on  the chip.   One  $I$  standard  is observed  before  the
observation of a strip and  one $J$ and $K_s$ standard afterwards.  On
average  $6$  to  $8$ strips  per  night  are  observed.

Data on  the MCs  were taken during  observing seasons from  August to
March,  the first  centered  on December  1995,  and the  last one  on
December 1998. The two clouds, LMC  and SMC, were covered by $119$ and
$88$ strips, respectively.

\section{Data Reduction}
Data  reduction took  place in  two centers:  the Paris  Data Analysis
Center  (PDAC)  and the  Leiden  Data  Analysis  Center (LDAC).   PDAC
pre--processed  the  raw  data  and LDAC  extracted  and  parametrized
objects  ranging  from  point   sources  to  small  extended  sources.
The reduction at LDAC by the first author led to the detection
of  numerous technical  problems (astrometry  for example),  which had
escaped the checks  of the automatic pipeline.  Compared  to the first
relase  of DENIS  data (Epchtein  et al.  \cite{epal})  this catalogue
differs   in  terms   of  flags,   astrometric   reference  catalogue,
association   criteria  and  photometric   calibration  (use   of  the
overlapping  region between adjacent  strips). Besides,  our catalogue
covers a portion of the sky  not overlapping with the first DENIS data
release.  The  source list (table  7) and the  photometric information
(table 8) are  electronically available at CDS via  this catalogue. 
With further DENIS data releases data on a strip by strip basis,
therefore not merged into a single catalogue without treating 
 the overlapping regions in terms of photometry and astrometry,  
will also be available. This means that all the strips covering the same 
region of the sky regardless of their quality will be available; multiple 
entries for a single objects could then be retrieved.
We show  in section  3.2.3 the  consistency of  our calibration
within each cloud.

The  work  of Fouqu\'e  et  al.  \cite{fou}  focuses on  the  absolute
photometric  calibration of  the  DENIS data.  This calibration,  once
completed  after  the  termination  of  the survery,  might  induce  a
systematic shift on the photometry of the present catalogue.

\subsection{PDAC: Paris Data Analysis Center}
At  the  PDAC the  images  were
corrected for sensitivity differences and atmospheric and instrumental
effects.   Their  optical quality  was  judged  on  the basis  of  the
parameters  that  describe  the  point spread  function  (see  Sect. 
$3.1.2$).

\subsubsection{Flat and Bias}
The received intensity from  the target image also contains background
contribution  from  the   telescope  and  the  atmospheric  radiation.
Besides, the  sensitivity varies across  the array of the  camera.  The
true signal ($TS$) for the pixel ${i,j}$ is obtained from:
\begin{equation}
TS_{i,j} = I_{i,j}(t)-F_{i,j}\times b(t) - B_{i,j}\, ,
\end{equation}
where $I_{i,j}$ is the measured intensity after subtraction of dark currency, 
$F_{i,j}$ is
the flat field after dark subtraction, $b(t)$ is the background and $B_{i,j}$
is the bias level. 
The background is estimated per image, at time $t$, with
\begin{equation}
b(t)=\frac{\sum_{ij}^{N}[I_{i,j}(t)/F_{i,j}]}{N}\, ,
\label{back}
\end{equation}
where $N$ is the total number  of pixels per image.  At this stage we use the
flat field  and the  dark current values  estimated from  the previous
night are  used.  Points outside  the $3\sigma$ level are  rejected in
the sum.  The four $I$ quadrants are treated separately. This is also
done for each of the $9$ sub-images in the $J$ and $K_s$ bands.

As a  second step  we select low--background images  from the sunrise  
sequence ($180$ in  a normal strip)  with a  low background  value. 
To  identify and
avoid crowded fields  and fields affected by saturated stars
we  combine  measurements taken  during  different  nights.  We  then
determine the flat $F_{i,j}$ and  the bias $B_{i,j}$ by minimizing the
expression:
\begin{equation} 
\sum_{t=1}^{N} [I_{i,j} (t)- F_{i,j} b (t) - B_{i,j} ]^2\, , 
\label{flabia}
\end{equation}
where $N$  is the number  of selected images ($\leq 180$).   
In the third  step, we
applied the new values of the flat and the bias to the set of selected
images  to  obtain  a  new  estimate of  the  background  value,  more
appropriate to the particular night.  The quality of the determination
of  the parameters  involved is  improved  by iteration  of the  above
procedure.

The bias so far determined is  a mean value for the night.  Because it
varies during the  night, its value for a given  strip is estimated to
be:
\begin{equation}  
B_{i,j} =  \frac{\sum_{t=1}^{N}[I_{i,j} (t)- F_{i,j} b (t)]}{N}\, ,
\label{bias}
\end{equation}
where $N$ is the total number of images per strip ($180$).  After dark
subtraction,  the   bias  contains   only  the  contribution   of  the
instrumental and  atmospheric emission which  does not affect  the $I$
band,  but does  affect strongly  the $K_s$  band and  a higher
number of iterations is sometimes necessary.

The large  number of available flat/bias-images ($180$)  gives a quite
high degree of statistical  confidence to both determinations. This is
not  true in  case  of  calibration sequences  that  involve only  $8$
images. In  this case,  the bias determined  for the strip nearest in
time is applied.

\subsubsection{Point Spread Function}
The pixel  size of the $J$ and $K_s$ channels is $3\arcsec$ and the sampling is
$1\arcsec$ in both directions.  The  real width of any point source is
therefore potentially narrower than  the pixel ($3\arcsec$).  In terms
of signal processing the sources are not under-sampled, but the width of
the filter is broader than the  sampling.  To estimate the width of the
signal  the   convolution  of  the  signal  profile   (assumed  to  be
elliptical)  and the  pixel size  has  been taken  into account.   The
method of least squares has  been applied to the projection of sources
onto RA, DEC and diagonal axes (Borsenberger, \cite{bors}). In the $I$
and  the $J$  bands there  are enough  sources to  build a  model that
describes the behaviour of the  projected widths in each image.  In the
$K_s$  band  several  images   were  stacked  together  prior  to  the
determination.

We   refer  to  ht\-tp:\-//www\--denis.\-iap.\-fr/\-docs/\-te\-ne\-ri\-fe.\-html   for  more
details on the PDAC data reduction.

\subsection{LDAC: Leiden Data Analysis Center}
LDAC  extracts point sources from
the images delivered by PDAC. From these sources, it derives and
then  applies  astrometric and  photometric  calibration  to obtain  a
homogeneous  point   source  catalogue.   The   astrometric  reference
catalogue     is     the     USNO--A2.0     (Monet     \cite{monet}) 
that provides on average $100$ ``stars'' per DENIS image.
 
The photometric  DENIS standard stars belong  to different photometric
systems  of  which the  major  ones  are:  Landolt (\cite{lan}),  Graham
(\cite{gra}), Stobie et al.  (\cite  {sgr}) and Menzies et al. (\cite{mcbl})
in $I$; Casali \& Hawarden (\cite{cashaw}), Carter (\cite{car}) and Carter
\& Meadows  (\cite{carmea}) in $J$  and $K_s$. An  absolute calibration,
together  with a  definition of  DENIS photometric  bands is  given by
Fouqu\'e et al. (\cite{fou}).

\subsubsection{Source Extraction}
The first LDAC task is to  reduce the information from each image into
an object list. This is  done using the SExtractor program (Bertin and
Arnouts            \cite{emb})            version            $2.0.15$.

\subsubsection{Astrometric Calibration}
Positions  are determined  through pairing  information  among frames,
channels and  with the reference catalogue.   The astrometric solution
makes  use of  the fact  that each  map has  an area  of  overlap with
neighboring maps, and that objects in the overlapping region have been
observed many times.  The  projected position of the multiply observed
sources, in  terms of their  pixel positions, contains  information on
the  telescope   pointing  and  the  plate   deformations.  The  plate
deformation  is derived  through a  triangulation  technique, matching
bright  extracted  objects with  astrometric  reference objects.   The
resulting global  solution for each strip takes  into account possible
variations along the strip. The plate offsets are determined using all
but  the faintest  extracted  objects, matching  among channels  (wave
bands)  and in  overlap.  A  least  square fitting  technique is  then
applied to the functional  description of the detector deformation and
its variation to obtain the full  solution on the basis of the pairing
information.  Thereafter,  the celestial  position, its error  and the
geometric parameters of each object are calculated.

The standard  position accuracy  derived is RMS  $0.001$ arc  sec with
maximum  excursions of  $1.32$ arc  sec. This  error is in addition  
to the  RMS of  $0.3$ arc  sec of  the  astrometric reference
catalogue.

\subsubsection{Photometric Calibration}
\begin{figure}
\resizebox{\hsize}{!}{\includegraphics{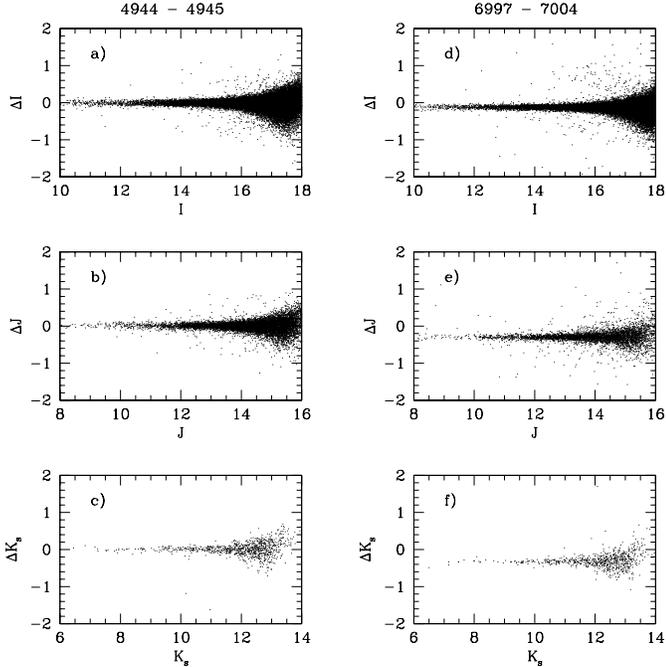}}
\caption{Magnitude differences of overlapping sources, between strip 4944 and strip 4945 observed under good photometric conditions (a, b, c) and between strips 6997 and 7004 (d, e, f), strip 7004 was observed under poor photometric conditions.}
\label{diff}
\end{figure}

Magnitudes are  estimated within a circular aperture  of $7\arcsec$ in
diameter after  a de--blending  process, that determines  which pixels
are within  the aperture,  and what fraction  they contribute  to each
individual  source  (Bertin \&  Arnouts,  \cite{emb}).  This  aperture
collects $95\%$ of the light when considering a seeing of $1.5\arcsec$
and the  pixel size  of $3\arcsec$ for  the infrared wave  bands.  For
homogeneity  we used the  same aperture  also for  the $I$  band.  The
source magnitude  ($m$) corresponding  to the wavelength  $\lambda$ is
defined as:
\begin{equation}
m_{\lambda} = -2.5\log(S_{\lambda}) + m_{\lambda0}\, ,
\label{mag}
\end{equation}
where $S_{\lambda}$  is the  observed flux and  $m_{\lambda0}$ defines
the zero--point  of the magnitude  scale at the  wavelength $\lambda$.
The  determination  of  the  instrumental quantity  $m_{\lambda0}$  to
correct the  stellar magnitude  for atmospheric effects  is done  on a
nightly basis.  First, standard star measurements are matched with the
information  stored  in  the  standard  star  catalogue.   Second  the
instrumental zero-point  ($m_{\lambda0}$) is  derived for each  of the
eight measurements  of the standard  star assuming a  fixed extinction
coefficient,  $\epsilon$  (Eq.  \ref{m_0}).   The  adopted  values  of
$\epsilon$ are $0.05$ for the $I$  band and $0.1$ for both the $J$ and
$K_s$  bands.   These values have been determined from the photometric
measurements  performed  during calibration  nights  (nights where  only
standard stars were observed).
\begin{equation}
m_{\lambda0} = 2.5\log(S_{\lambda}) + m_{\lambda \mathrm{ref}} + \epsilon \times z\,
\label{m_0}
\end{equation}
$m_{\mathrm{ref}}$  is the  magnitude of  the standard  star  from the
standard star catalogue and $S_{\lambda}$ is the flux as measured at a
given air mass ($z$). Standard  stars were selected near the 
airmass limit of the strips and to be  roughly of the  same spectral
type;  this simplifies  the  Taylor expression  used  to describe  the
extinction law  because colour  terms (Guglielmo et  al. \cite{gugli})
and the non--linear terms are of minor importance; in the infrared the
dependence  of  the   extinction  on  $z$  is almost  linear for $z<2$.   
In principle, both  $m_{\lambda0}$ and $\epsilon $  can be determined 
simultaneously and the non--linear terms can be incorporated as well 
if a sufficient number of star measurements are available, 
 but for a single  night there are not enough, in fact the 
use of the approximated law (Eq. \ref{m_0}) gives a systematic offset between
the  magnitude of  the source in the overlap of two strips of  
comparable, but different,
 photometric conditions.  After a considerable investigation
it turned  out that  this offset  could be greatly  reduced if  a fixed
extinction coefficient is used.
Some differences are left  when  the  observations have  been
performed in different photometric conditions or when too few standard
star measurements were done.  Fig. \ref{diff}  shows the computed
differences  between the  magnitudes of  the sources  detected  in the
overlapping region of two strips observed under comparable photometric
conditions ((a), (b),  (c)) and of two strips  observed with different
photometric  conditions ((d),  (e), (f))  in  the $I$,  $J$ and  $K_s$
bands,  respectively.  Faint  sources give rise to a larger  dispersion.  The
systematic  shift  is  clearly visible  in  Fig.  \ref{diff}d--f.

The final  nightly value of  $m_{\lambda0}$, for  each wave
band, is  calculated by averaging the single  determinations for each
standard star  and among all  the standard stars observed  during that
night, after removal of  flagged (Sect.  4) measurements (this reduces
on average the number of measurements per star from $8$ to $6$).  The
flagged measurements  have a  non-zero value for  at least one  of the
types  of flag considered  in the  pipeline reduction.  Only standards
fainter  than  $I=10.5$, $J=8.0$  and  $K_s=6.5$  mag  are used.   The
instrumental $m_{\lambda0}$ and its  standard deviation are listed for
each  strip in the  quality table  (table \ref{quatab}).   Mean values
($\pm 1\sigma$) are: $23.42\pm  0.07$ ($I$), $21.11\pm 0.13$ ($J$) and
$19.12\pm 0.16$ ($K_s$).

Using  the  overlapping regions  of  adjacent  strips to correct 
for remaining differences we performed  a
 general photometric  calibration, separately for the LMC  and for the
 SMC.   We  calculated the  magnitude  difference of  cross-identified
 sources between two adjacent strips of sources detected in three wave
 bands.  The histogram of these  differences in magnitude shows when a
 systematic shift is present between the two strips (Fig. \ref{diff}).
 In only  a few cases is  the average magnitude affected  by more than
 0.1 mag.   If necessary we applied a  systematic shift (table
 \ref{quatab}).   Experience   showed  that  if  a   strip  is  poorly
 calibrated the magnitude difference  in the overlap with the previous
 strip has a sign opposite to the difference found in the overlap with
 the  next strip. Note that observing a strip in good photometric 
 conditions but having too few standard star measurements to perform the
 calibration may induce alone this offset; the equal number of detected objects
 as a function of magnitude per  band 
 in both strips indicates, as just mentioned, a minor difference in the
 photometric   conditions
   under  which   each  strip   was  observed, 
 increasing the confidence we  place in the correction procedure. Only
 $9$ strips  out of $108$ for  LMC observations and $3$  strips out of
 $81$ for SMC observations show this behaviour.  The maximum observed
 shift amounts to  2.45 mag in the $J$  band. Sources with corrected  
magnitude are easily recognized from their strip number 
associated to each detected band (Table \ref{soutab}). 
Table \ref{quatab} reports the amount of the 
applied shift as a function of strip number.
   In
 some cases, the difference shows a dependence on declination, but the
 effect  on the  averaged magnitude,  in  the area  of the  Magellanic
 Clouds, is not significant (less than $0.1$ mag), and can be ignored.
 The internal statistical  RMS error is between $0.001$  and $0.4$ mag
 at  the  detection limit,  faint  sources  have  larger errors.   For
 completeness we included in  the catalogue sources detected above and
 below   the  reference   saturation  and   detection   limits,  their
 photometric errors (larger than $0.4$ mag) show the confidence of the
 detection.  The  standard deviation on  $m_{\lambda0}$ is in  most of
 the  strips  below $0.05$  mag,  but  spreads  from $0.01$  to  $0.2$
 mag. Larger  values are  detected in the  strips where  a photometric
 shift  was also  applied, therefore  the resulting  accuracy  is, for
 these few cases,  not better than $0.1$ mag.  In  all other cases the
 resulting accuracy has an RMS error better than $0.05$ mag.

\subsubsection{Association}
All  extracted objects  are matched  on the  basis of  their geometrical
information assuming  an elliptical  shape (RA, DEC,  $a$: semi--major
axis, $b$:  semi--minor axis, $\theta$: inclination  angle) within one
wave  band,  among the  three  wave bands  within  a  strip and  among
different  strips.   The  geometrical   parameters  of  each  object  are
evaluated at the $3\sigma$ level of the row image; $a$ and $b$ are the
second order moments of the  pixel distribution within the size of the
photometric aperture.   Typical values are  $1.8\arcsec$, $1.0\arcsec$
and $0.5\arcsec$  for the $I$,  $J$ and $K_s$ band,  respectively, 
differences among the three wave bands mainly depend on the 
differences 
in sensitivity; the second order moments characterize the PSF. The
effective area  used during the  association procedure is  $1.5$ times
(tolerance) the area defined by  the $a$ and $b$ values of 
both object,  when the association is  performed within each  band of a
strip. Sources  previously   de--blended  are  not 
associated.  When  the association is  done among different  bands the
tolerance value increases to $2.5$.

We associate  two objects when the  center position of one  of them is
within the bounds  of the ellipse of the other, even  if the center of
the second  is outside the ellipse  of the first one,  and vice versa.
For the coordinates,  we always used a weighted  average (based on the
signal to noise  ratio and detection conditions  as derived from
the source  extraction program and the  astrometric calibration).  For
the magnitudes we decided not to average or to combine magnitudes from
different  epochs (strips) because  of the  possible variability  of a
large fraction of the detected objects.  Objects associated within the
same strip  are given with the average of
the magnitudes.   When the association involves  overlapping strips we
distinguish the following cases: (1) for objects detected in all three
wave bands in  both strips we choose the entry from  the strip with the
lowest value  of $\sum_{i=1}^{N}\sqrt{a_ib_i}$, where  N is the  number of
sources  detected in  the  overlap;  (2) for  objects  detected in  an
unequal number of  wave bands, we chose the entry  from the strip with
the highest number of detected wave bands; (3) for objects detected in
two different  wave bands we choose  the entry from the  strip with the
lowest $\sum_{i=1}^{N}\sqrt{a_ib_i}$,  including the third  magnitude from
the  other  strip.  When  the  strip  numbers  of  the
detected wave  bands differ the observations refer to different  epochs.  The
criteria  given  conserve  the  major  property  of  the  DENIS  data:
simultaneousness.

We  refer  to  
ftp.\-strw.\-lei\-den\-univ.\-nl /pub/\-ldac/\-pi\-pe\-li\-ne.\-ps  for  more
details on the LDAC data reduction.

\section{Data Quality}
For  each   image  PDAC  flags   problems  of  different   kinds  (see
table \ref{imaflag}).  The  flags are used  by LDAC to  identify image
defects and consequently flag the extracted objects if necessary.

  \begin{table}
      \caption[]{Image Flags, mainly PDAC information}
         \label{imaflag}

      \[
         \begin{array}{rl}
            \hline
            \noalign{\smallskip}
            \mathrm{Value}  & \mathrm{Description} \\
            \noalign{\smallskip}
            \hline
            \noalign{\smallskip}
1\,\,\, & \mathrm{Regression\,failed\,for\,this\,pixel}\\
2\,\,\, & \mathrm{Flat\,derived\,is\,less\,than\,0.01\,(dead\,pixels)}\\
3\,\,\, & \mathrm{Abnormally\,high\,flat}\\
4\,\,\, & \mathrm{Abnormally\,high\,residual\,after\,adjustment\,(5\sigma)}\\
5\,\,\, & \mathrm{Bias\,more\,than\,5\sigma\,of\,mean}\\
6\,\,\, & \mathrm{Flat\,more\,than\,5\sigma\,of\,mean}\\
9\,\,\, & \mathrm{Insufficient\,number\,of\,pixels\,left\,after\,iterative\,clipping}\\
16\,\,\, & \mathrm{Flat\,<\,0.7\,and/or\,flux\,<\,-200\,ADU}\\
32\,\,\, & \mathrm{Flat\,>\,1.3\,and/or\,flux\,>\,49500\,ADU}\\ 
           \noalign{\smallskip}
            \hline
         \end{array}
      \]

\mbox{Note that these flags do not  have binary exclusive values.}  
   \end{table}

During  the  source  extraction   process,  LDAC  produces  more  flag
information  (tables \ref{extflag}  and \ref{artifact}).  Artifact
flags are not  present in the output parameter  list because they have
been used  as a  primary selection criteria  to filter  the catalogue;
most of  the cosmic rays, glitches  and optical ghosts  will have been
eliminated.

  \begin{table}
      \caption[]{Extraction Flags (from SExtractor), all LDAC information}
         \label{extflag}

      \[
         \begin{array}{rl}
            \hline
            \noalign{\smallskip}
            Value  & Description \\
            \noalign{\smallskip}
            \hline
            \noalign{\smallskip}
1\,\,\,   & \mathrm{Bright\,neighbours\,or\,bad\,pixels}\\
    & \mathrm{(more\,than\,10\%\,of\,the\,integrated\,area\,affected)}\\
2\,\,\,   & \mathrm{The\,object\,was\,originally\,blended\,with\,another}\\
4\,\,\,   & \mathrm{At\,least\,one\,pixel\,of\,the\,object\,is\,saturated\,}\\
8\,\,\,   & \mathrm{The\,object\,is\,truncated\,(too\,close\,to\,image\,boundary)}\\
16\,\,\,  & \mathrm{Objects\,aperture\,data\,are\,incomplete\,or\,corrupted}\\
32\,\,\,  & \mathrm{Objects\,isophotal\,data\,are\,incomplete\,or\,corrupted}\\
64\,\,\,  & \mathrm{A\,memory\,overflow\,occured\,during\,deblending}\\
128\,\,\, & \mathrm{A\,memory\,overflow\,occured\,during\,extraction}\\
            \noalign{\smallskip}
            \hline
         \end{array}
      \]

   \end{table}

  \begin{table}
      \caption{Artifact Flags, all LDAC information}
         \label{artifact}

      \[
         \begin{array}{cl}
            \hline
            \noalign{\smallskip}
            Value  & Description \\
            \noalign{\smallskip}
            \hline
            \noalign{\smallskip}
1 & \mathrm{The\,object\,is\,probably\,a\,glitch}\\
2 & \mathrm{The\,object\,is\,probably\,a\,ghost}\\
4 & \mathrm{The\,object\,is\,saturated\,(}I\mathrm{\,band)}\\
            \noalign{\smallskip}
            \hline
         \end{array}
      \]
   \end{table}

When dust is present on the mirrors of the telescope and on the lenses
of the  instrument spurious objects  are created. Most are  bright and
easily recognizable.  During the pipeline reduction photometric fluxes
are     calculated     in     $7\arcsec$    and     in     $15\arcsec$
apertures. ``Dusty--like  objects'' give a  negative flux in  the larger
aperture (and  its value  is set to  $99$).
 In  their proximity the
flat value for the pixels is dominated by their continuous presence in
all images  along the strip, therefore we end up with  an area
with negative  flux next to  the 'dusty--like objects'.  This  area is
not always  in the same position  because of bending  of the telescope
during  the  observation of  a  strip.   To  eliminate these  spurious
detections we required that  both aperture fluxes were positive.  This
selection also allowed the removal of glitches not previously flagged,
sources too  close to  the image  borders or too  close to  broad dead
pixel regions  and dummy sources with photometric  errors greater than
$0.2$ mag.

\begin{figure}
\resizebox{\hsize}{!}{\includegraphics{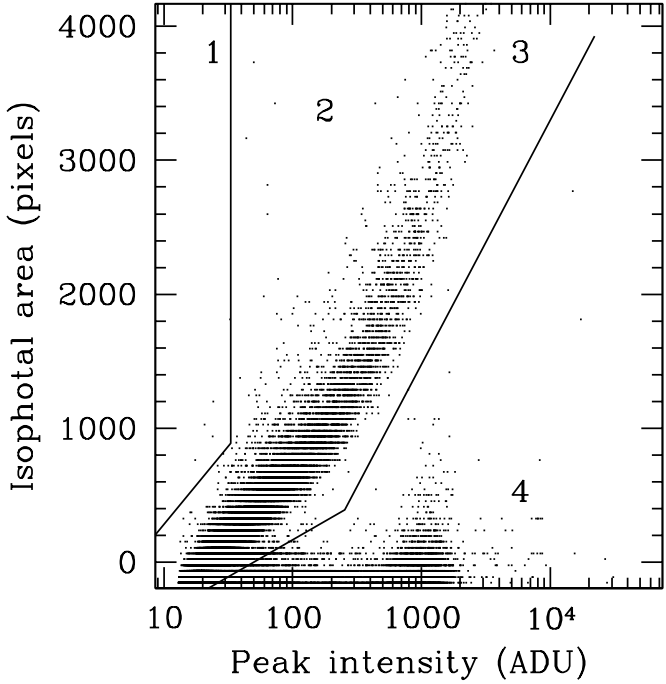}}
\caption{Object isophotal area (Area) versus maximum peak intensity (I--peak) for strip 5029 in the $J$ band.}
\label{glitch}
\end{figure}

An  additional filtering  criterion is  based  on the  diagram of  the
isophotal area of  one object at the $1\sigma$ level  of the raw image
(Isophotal area--pixels)  versus the peak  intensity (ln(MaxVal), Peak
intensity--ADU); see Fig. \ref{glitch}. Area (3) of point sources
(stars)  is  clearly  identified:  the objects  have  a  Gaussian
intensity  energy distribution. Area (2): galaxies are  extended  objects and,
relative  to  stars,  their   area  increases  faster  for  increasing
intensity--ADU.   The broadening of the  locus of stars  is due to
the variation  of the PSF over  the field and of  the seeing.  Areas
(1) and (4) contain cosmic rays and electronic glitches and are easily
distinguishable.  We accepted only sources in areas (2) and (3); the
same  cut between stars/galaxies/glitches--cosmic  rays was  applied to
all strips.

Finally we eliminated sources for which the object PSF could not match 
the instrumental PSF. This led to the loss of 
a few percent of $K_s$ detections;
 this effect does not depend on the source brightness and arises as a 
consequence of image defects.

Filtering  based on  the flags,  on dust  on the  detector and  on the
previous diagram  were applied before we made the cross  identification 
between the different wave bands.

\subsection{Completeness}

A  few strips  had  to be rejected during the reduction 
phase because of poor  quality. These strips have been re--observed,
but the data reduction is  not yet started and the strips
have not been included in  the catalogue.  Table
 \ref{missregs} lists  the right ascensions of  the absent strips
(each covers $1^m20^s$ in  RA).  Considering the overlap with adjacent
strips  we have  missed  $7.7$ \%  and $6.3$  \%  of the  LMC and  SMC
regions, respectively.

  \begin{table}
     \caption[]{Right Ascension of the absent strips, each $1^m20^s$ wide}
         \label{missregs}

      \[
         \begin{array}{lcclc}
            \hline
            \noalign{\smallskip}
        \mathrm{Cloud}&\mathrm{Ra (2000)}& &\mathrm{Cloud}&\mathrm{Ra (2000)}\\
            \noalign{\smallskip}
            \hline
            \noalign{\smallskip}
	    \mathrm{LMC} & 4^h 17^m 20^s & & \mathrm{LMC} & 6^h 00^m 00^s \\
	    \mathrm{LMC} & 4^h 45^m 20^s & & \mathrm{LMC} & 6^h 40^m 00^s \\
	    \mathrm{LMC} & 4^h 46^m 40^s & & \mathrm{SMC} & 0^h 04^m 00^s \\
	    \mathrm{LMC} & 4^h 50^m 40^s & & \mathrm{SMC} & 0^h 26^m 40^s \\
            \mathrm{LMC} & 5^h 01^m 20^s & & \mathrm{SMC} & 0^h 37^m 20^s \\
	    \mathrm{LMC} & 5^h 30^m 40^s & & \mathrm{SMC} & 0^h 49^m 20^s \\
	    \mathrm{LMC} & 5^h 32^m 00^s & & \mathrm{SMC} & 1^h 18^m 40^s \\
	    \mathrm{LMC} & 5^h 34^m 40^s & & \mathrm{SMC} & 1^h 49^m 20^s \\
	    \mathrm{LMC} & 5^h 38^m 40^s & & \mathrm{SMC} & 1^h 56^m 00^s \\
            \noalign{\smallskip}
            \hline
         \end{array}
      \]
   \end{table}
 
We now consider the  completeness of  the  catalogue under two different
aspects: completeness of objects detected in only two wave bands or in
all three wave bands.

\begin{figure}
\resizebox{\hsize}{!}{\includegraphics{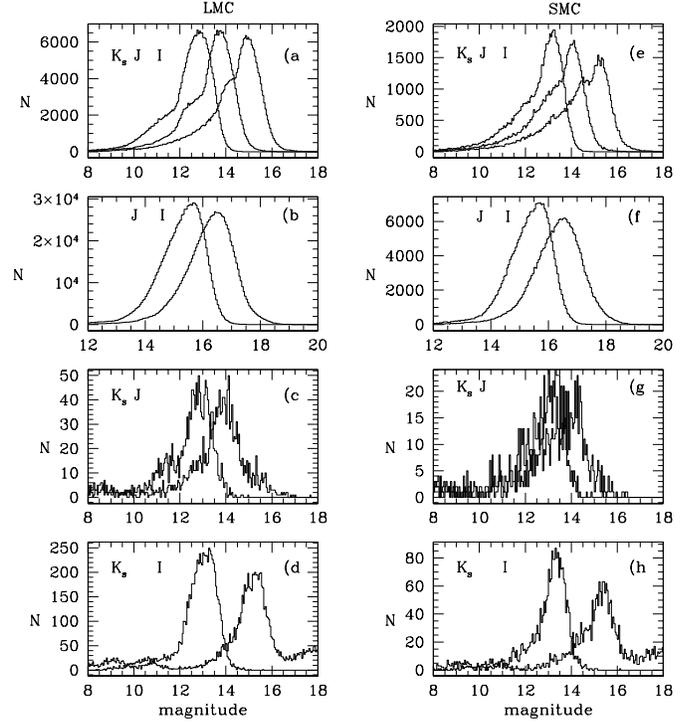}}
\caption{Differential star counts of sources in 0.05 mag bins for the whole catalogue, with detections in three or two bands. From left to right: LMC and SMC. From top to bottom: sources detected in three wave bands, sources detected only in $I$ and $J$, sources detected only in $J$ and $K_s$ and sources detected only in $I$ and $K_s$. Note that the tip of the RGB is very nicely detected.}
\label{comp}
\end{figure}

Fig. \ref{comp} displays histograms of the number of sources in the
catalogue in $0.05$ mag  bins. Fig. \ref{comp}a--d
refer  to  the  LMC  and  Fig. \ref{comp}e--h  to  the
SMC. Table \ref{comptab} contains the  magnitude of the maxima in the
various histograms.

A  full  discussion  of  these  histograms  will  be  given  elsewhere
(M.R.  Cioni,  H.J. Habing,  M.  Messino,  in  preparation). We  limit
ourselves to a few comments.

(1) Comparing Fig. \ref{comp}a  and Fig. \ref{comp}b, and  similarly
    Fig. \ref{comp}e  and Fig. \ref{comp}f  suggests  that (\ref{comp}b)  and
    (\ref{comp}f)  contain  sources  similar  to  (\ref{comp}a)  and
    (\ref{comp}e),  
    but they are below the detection limit in the $K_s$ band. 
    Fig. \ref{comp}b and \ref{comp}f contain many more sources than 
    Fig. \ref{comp}a and \ref{comp}e, respectively.

(2) The $I$ and  $K_s$ histograms of
    (\ref{comp}d)  and (\ref{comp}h)  are approximately  scaled down
    versions  of the $I$  and $K_s$  histograms in  (\ref{comp}a) and
    (\ref{comp}e). This suggests that they contain the same kind of sources, 
    and that the sources in (\ref{comp}d) and (\ref{comp}h) have not been 
    detected in the $I$ band, i.e. the detection rate in the $J$ band is 
    never $100$\%, although it will be very close.

(3) The  nature of  the sources  in (\ref{comp}c)  and (\ref{comp}g)
    remain unspecified for the moment.

(4) The   magnitudes  of  the   maximum  count   as  given   in  table
     \ref{comptab} show  that the magnitudes referring to  the SMC are
    about $0.25$ mag  fainter --  this reflects the  larger distance  to the
    SMC. This conclusion is not true for the counts of sources detected only in
    $I$ and $J$. These counts may contain a large foreground component.

\begin{figure}
\resizebox{\hsize}{!}{\includegraphics{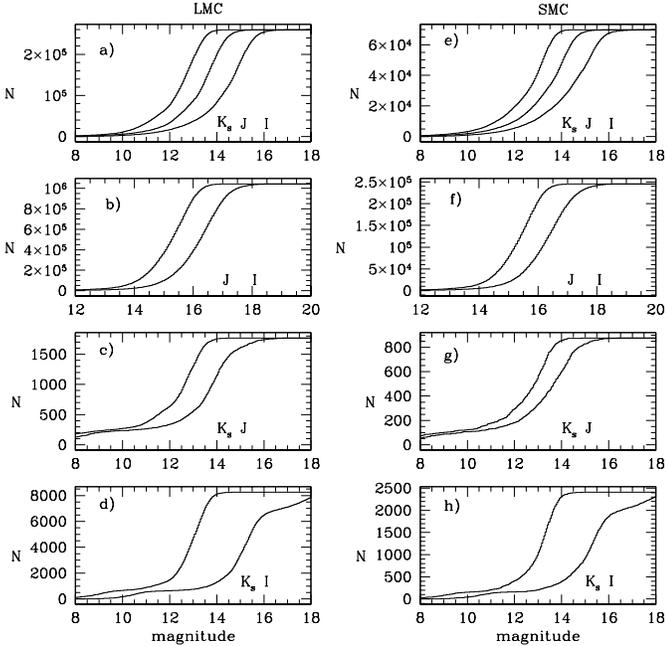}}
\caption{Cumulative star counts of sources in 0.05 mag bins for the whole catalogue, with detections in three or two bands. From left to right: LMC and SMC. From top to bottom: sources detected in three wave bands, sources detected only  in $I$ and $J$, sources detected only in $J$ and $K_s$ and sources detected only in $I$ and $K_s$.}
\label{cum}
\end{figure}

Fig. \ref{cum} displays the cumulative distributions of the sources in the 
catalogue.

  \begin{table}
     \caption[]{Magnitude of maximum count (see Fig. 3 and text)}
         \label{comptab}

      \[
         \begin{array}{lcccclcccc}
            \hline
            \noalign{\smallskip}
  \mathrm{Cloud}&I&J&K_s&\mathrm{Fig.}&\mathrm{Cloud}&I&J&K_s&\mathrm{Fig.}\\
            \noalign{\smallskip}
            \hline
            \noalign{\smallskip}
\mathrm{LMC}&15.00&13.75&12.75&3\mathrm{a}&\mathrm{SMC}&15.25&14.00&13.25&3\mathrm{e}\\
\mathrm{LMC}&16.50&15.25&     &3\mathrm{b}&\mathrm{SMC}&16.50&15.75&     &3\mathrm{f}\\
\mathrm{LMC}&     &14.00&12.75&3\mathrm{c}&\mathrm{SMC}&     &14.25&13.25&3\mathrm{g}\\
\mathrm{LMC}&15.00&     &13.00&3\mathrm{d}&\mathrm{SMC}&15.50&     &13.25&3\mathrm{h}\\
            \noalign{\smallskip}
            \hline
         \end{array}
      \]
   \end{table}
 
From  the  overlap of  adjacent  strips, in  the  same  wave band,  we
estimate a  $5$\% difference in  the number of detected  sources. This
difference  is partly  due to regions of
insensitive  pixels on  the frame  borders, especially  in  the $K_s$
band.

\subsection{Galactic Foreground Sources}

Galactic  sources in  the foreground  have not  been removed  from the
catalogue. Therefore, we now discuss the probability that  any given source
belongs to the Magellanic Clouds or to the Milky Way Galaxy.

\begin{figure}
\resizebox{\hsize}{!}{\includegraphics{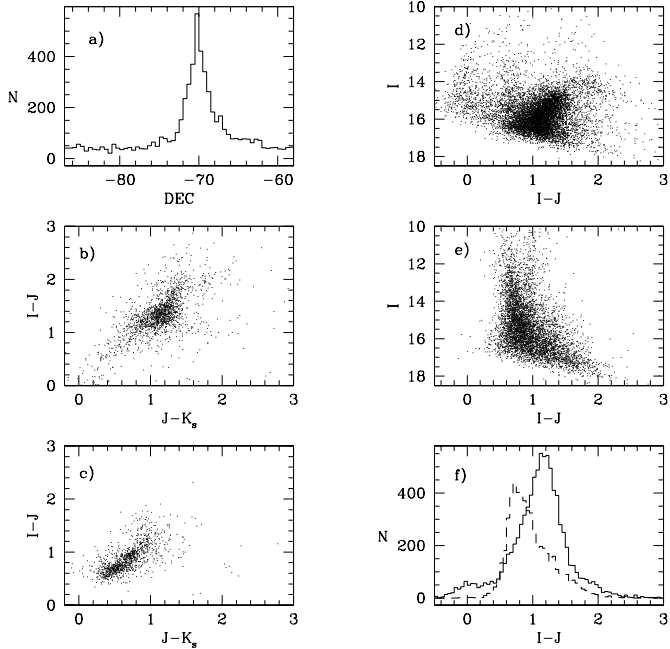}}
\caption{Sources detected towards the LMC in strip 5068. (a), (b) and (c) refer to sources detected in three wave bands and (d), (e) and (f) also include sources detected only in $I$ and $J$. In detail: (a) histogram of detected sources versus declination in bins of $0.5$ degrees, (b) colour--colour diagram of sources with $-69\degr>\delta>-71\degr$, (c) colour--colour diagram of sources with $\delta>-61$ and $\delta<-77$, (d) colour--magnitude diagram of sources with $-69\degr>\delta>-71\degr$, (e) colour--magnitude diagram of sources with $\delta>-61$ and $\delta<-77$, (f) continuous line: differential star counts in bins of $0.05$ mag of the same sources in (5d), dashed line: the same for sources in (5e).}
\label{gal}
\end{figure}

Fig. \ref{gal}a  shows that  the count of  sources detected  in all
three wave bands has a strong  maximum inside the LMC area
i.e. $-69\degr>\delta>-71\degr$. Outside of
this area  the count falls down to  a plateau at an  average value of
$50$ sources per $0.5$ degrees in declination; this plateau represents
the foreground contribution.

In  Fig.  \ref{gal}b we  show  the  colour--colour  diagram of  all
sources within the peak area of the  LMC, and in Fig. \ref{gal}c for all
sources outside  of the LMC.  The foreground sources  in (\ref{gal}c)
are probably ordinary dwarf stars  and red giants, for which we expect
colours  ($0.5$,$0.5$)  and  ($1.0$,$1.0$), respectively  (Bessell  \&
Brett \cite{besbret}).  The area outside  the LMC is about  $7$ times
the  area used  in Fig. \ref{gal}b  and this explains  why the
total number  of objects within  ($J-K_s<1$) and ($I-J<1$) is
much  larger in Fig. \ref{gal}c  than in Fig. \ref{gal}b: the
fraction  of foreground sources  in Fig.  \ref{gal}b is  very small
indeed.

Fig. \ref{gal}d, (e) and (f) refer  to sources detected in three wave bands 
plus sources detected only in $I$ and
$J$. The  comparison between Fig. \ref{gal}d (in the LMC) 
and Fig. \ref{gal}e (outside the LMC) shows
again what sources  may be galactic and what  sources are not. Sources
in Fig. \ref{gal}d with $I<16$ and $I-J>1.2$ are almost all
LMC objects. The same is true  for sources with $I-J<0.4$; these
are probably early type main--sequence  stars in the LMC. Sources with
$I>16$  and $I-J>1$  are  foreground  objects.

Fig. \ref{gal}f shows  the  histogram obtained  by  adding up  all
sources in  Fig. \ref{gal}d (full drawn  line) and in  Fig. \ref{gal}e (dashed
line) irrespective of the value of $I$. The difference in distribution
of points between Fig. \ref{gal}d and Fig. \ref{gal}e is obvious.
From all strips  and all colours we conclude  that, on average, $30$\%
of  the sources in the catalogue belong to the Galaxy rather than to 
the Magellanic Clouds. See also Cioni et al. (\cite{cioni}) 
for the separation of foreground and Magellanic stars within DENIS data.
A more  elaborate discussion will  be presented later (M-R.  Cioni and
H.J. Habing, in preparation).

\subsection{Confusion}
When  the source density  is too  high sources  will blend  with other
sources, a process  usually called confusion. A critical  value is $1$
source per $50$ detection elements (IRAS explanatory supplement, vol. 1,
VIII--4):  if   the  source   density  is  higher   confusion  becomes
statistically probable. The typical size of a detected source does not
exceed $2\arcsec$; see Sect. 3.2.4.

Fig. \ref{mapl} and Fig. \ref{maps} 
contain contour diagrams of source density in bins of constant right 
ascension. 
The maximum values is $500$ sources in $0.25\times 0.1$ square degrees in 
the LMC at $\delta = -70.5$ which implies  $1$ source per $200$ 
arcsec$^2$. This is well below the confusion limit.

Note that the confusion is not set by the size of the photometric aperture 
because the area of the aperture is independent of the detection process.
Within the aperture there might be two de--blended sources, each pixel belongs
to one or the other source or is shared between the two; the size of the 
aperture represents the contour limit where this pixel association process has 
to stop.

\section{Contents of the catalogue}
\begin{figure*} 
\resizebox{\hsize}{!}{\includegraphics{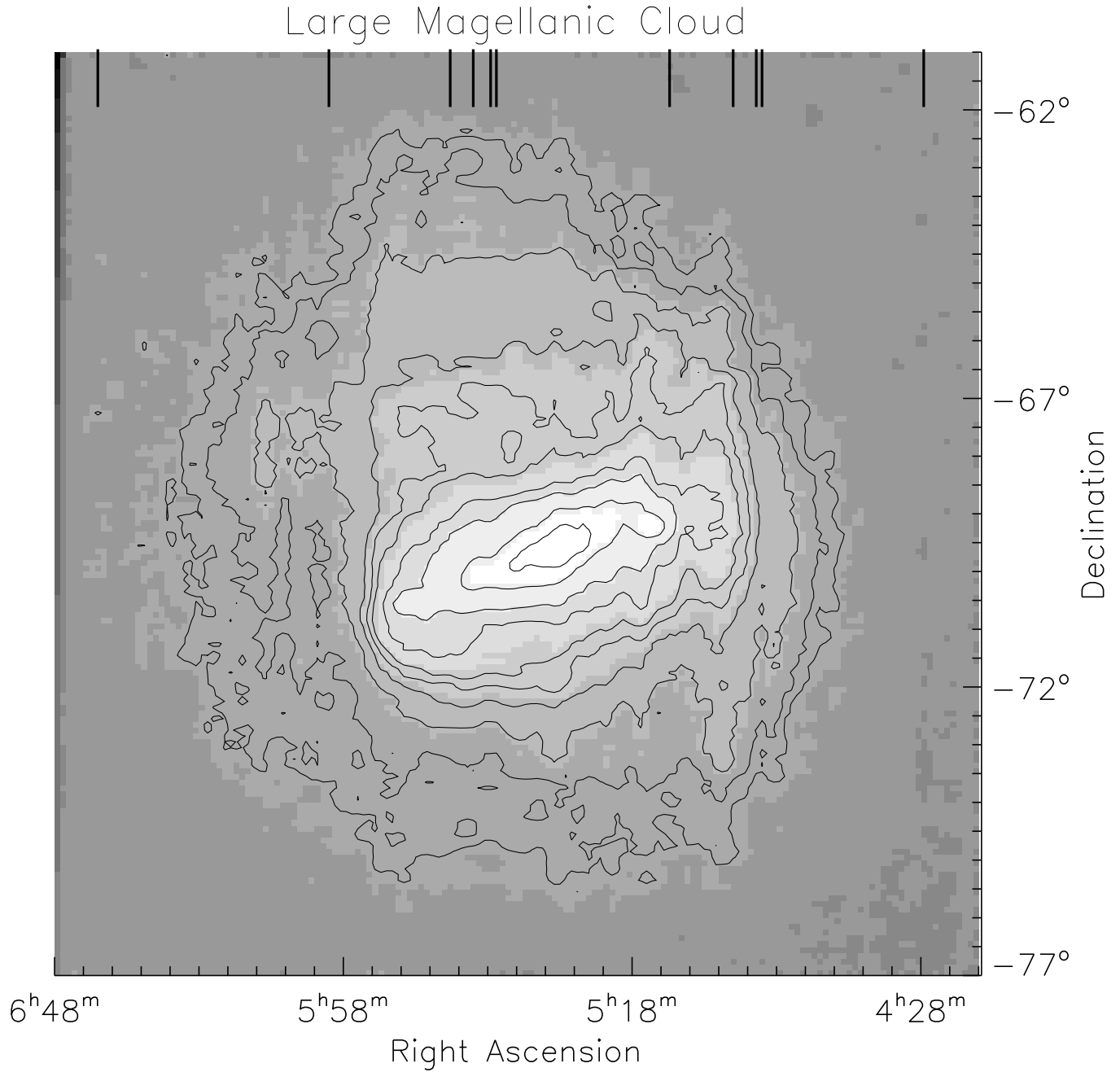}}
\caption{Density map towards the LMC per $0.25\times 0.1$ square degrees. Contours are at 1, 10, 25, 50, 75, 100, 150, 200, 300, 400, 500. Ticks on the upper horizontal axis indicate the position of the missing strips. The structure indicate the presence of a bar and spiral arms.}
\label{mapl}
\end{figure*}

\begin{figure*}
\resizebox{\hsize}{!}{\includegraphics{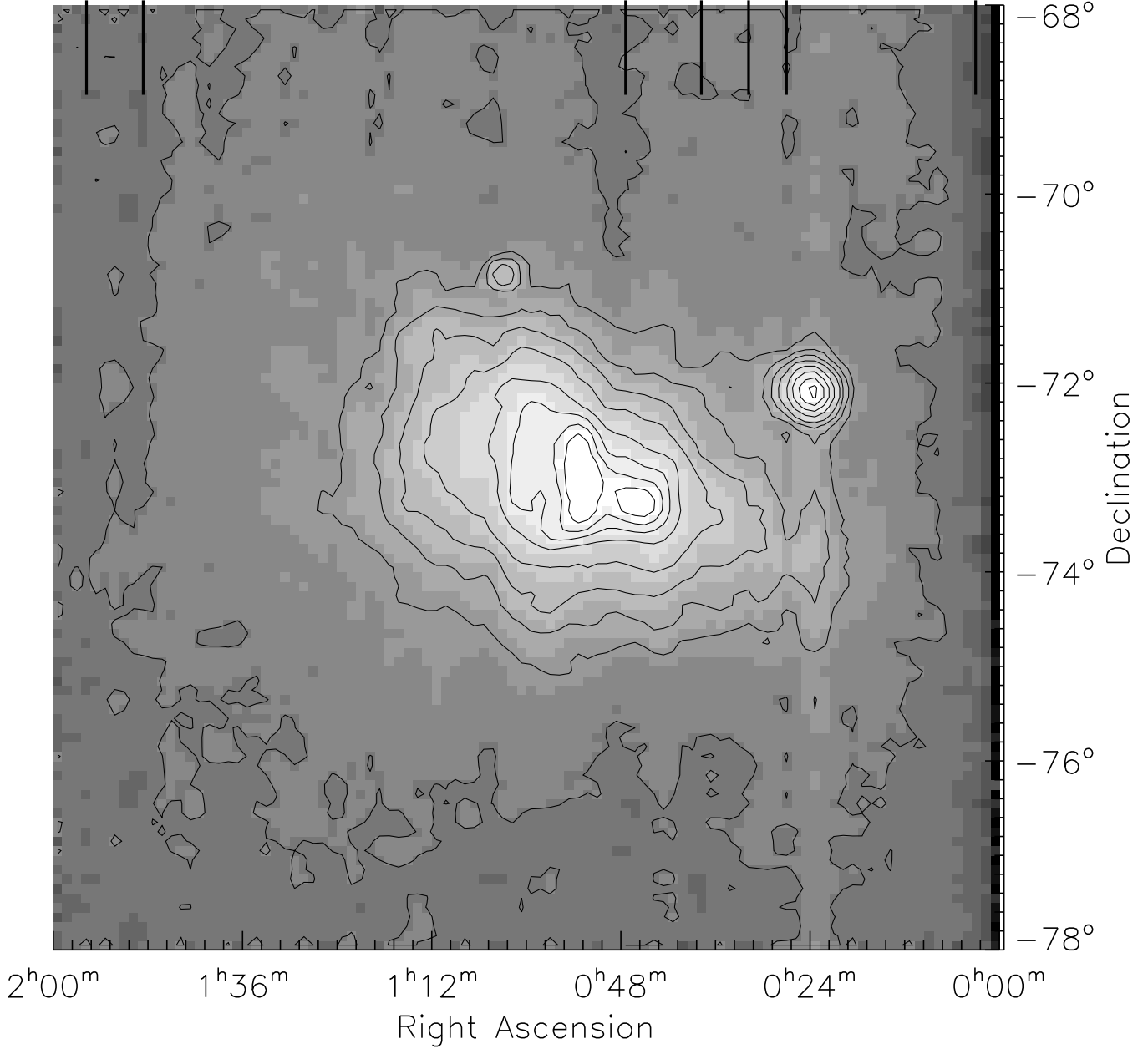}}
\caption{Density map towards the SMC per $0.3\times 0.1$ square degrees. Contours are at 18, 35, 50, 70, 100, 150, 200, 250, 300. Ticks on the upper horizontal axis indicate the position of the missing strips. The concentration of objects on the left side of the SMC structure is due to the galactic globlular cluster 47 Tuc.}
\label{maps}
\end{figure*}

The  present  version of  the  DENIS  point  source catalogue  of  the
Magellanic Clouds contains sources detected in at least two wave bands
within  the area  $4^h  08^m 00^s<\alpha< 6^h 46^m  40^s$,
$-61\degr>\delta>-77\degr$ for  the LMC and  $0^h 05^m 20^s<\alpha<2^h 00^m  00^s$, $-68\degr>\delta>-78\degr$
for  the  SMC  of  which  the  source  density  is  shown  in  Fig.
 \ref{mapl} and Fig. \ref{maps},  respetively.  The fraction of non--real
objects is  negligible as most glitches  are present only  in one wave
band.

Fig.  \ref{mapl} and  Fig. \ref{maps}  show the  density  maps of  all
detected sources towards the LMC and the SMC, respectively. The strips
not yet included are explicitely
indicated in the upper horizontal axis.
Fig. \ref{mapl} contains about $1\,300\,000$     sources and  Fig.
 \ref{maps} contains about $300\,000$.  Table  \ref{num} reports the 
 approximate number of sources detected  in three or two  wave bands.  

The two  parts of the
 catalogue  are  ordered by  increasing  RA.   Two  tables define  the
 meaning  of the  columns  in each  part:  a table  that contains  the
 detected sources (table \ref{soutab})  and a table that describes the
 quality   of   the   detections   on   a   strip   by   strip   basis
 (table \ref{quatab}).

  \begin{table}
     \caption[]{Detections}
         \label{num}

      \[
         \begin{array}{lrr}
            \hline
            \noalign{\smallskip}
          \mathrm{Wavebands}&\mathrm{SMC}&\mathrm{LMC}\\
            \noalign{\smallskip}
            \hline
            \noalign{\smallskip}
	    I+J+K_s &  70000 &  260000 \\
	    I+J     & 240000 & 1050000 \\
	    J+K_s   &    800 &    1800 \\
	    I+K_s   &   2300 &    8000 \\
            \noalign{\smallskip}
            \hline
         \end{array}
      \]
   \end{table}

\subsection{Nomenclature}
\subsubsection{Data table (table \ref{soutab})}
The star name is composed by  the acronym DCMC (DENIS Catalogue of the
Magellanic Clouds) and  by the coordinates of the  source at the epoch
2000  (\textit{Columns 1 and  2}) following  the IAU  convention 
`(Dubois et al. \cite{dub}).      For     example,     a     star  with
$(\alpha,\delta)=(1^h25^m20.15^s,-73\degr30\arcmin15.67\arcsec)$   would
have the designation DCMC $J012520.15-733015.6$; right ascension is
truncated at the second decimal  and the declination at the first.
\textit{Columns 3,  4, 5, 6, 7  and 8} give the  right ascension ($h$,
$m$, $s$)  and the declination ($\degr$, $\arcmin$,  $\arcsec$) at the
epoch J2000.  \textit{Column 9} gives  the positional error, i.e. 
the     statistical    error     calculated     during    the     data
processing. \textit{Columns  10 and 11} give the  pixel coordinates in
the  corresponding  image (\textit{Column  12})  of the  corresponding
strip  (\textit{Column 13}),  in the  $I$  band, where  the source  is
detected. The same quantities are  given in \textit{Columns 18, 19, 20
and 21} for the $J$ band and in \textit{Columns 26, 27, 28 and 29} for
the  $K_s$ band.   \textit{Columns  14,  22 and  30}  give the  source
magnitudes  and \textit{Columns  15, 23  and 31}  give  the associated
statistical  errors in  the $I$,  $J$ and  $K_s$  bands, respectively.
\textit{Columns   16,   24   and   32}  give   the   extraction   flag
(table \ref{extflag}) in the  three wave bands.  \textit{Columns 17, 25
and  33} give  the  image  flag (table \ref{imaflag})  in the  three
wave bands. \textit{Columns 34 and 35} give the $B$ and $R$ 
magnitudes from the cross-identification with the USNO--A2.0 catalogue.

\subsubsection{Strip quality table (table \ref{quatab})}
\textit{Column 1} gives the  strip number. \textit{Column 2} gives the
date of  observation of the given  strip. \textit{Columns 3  and 4}
give the  nightly zero--point and its standard deviation for the $I$ band.
The same quantities are given in \textit{Columns 5 and 6} for the $J$ band
and in \textit{Columns 7 and 8} for the $K_s$ band.
\textit{Column 9}  notes
information peculiar to the strip in question, for example 
the photometric shift (Sect. $3.2.3$).

  \begin{table*}
      \caption[]{Data table: detected sources}
         \label{soutab}

      \[
         \begin{array}{rlll}
            \hline
            \noalign{\smallskip}
            \mathrm{Col.}&\mathrm{Name}&\mathrm{Description}&\mathrm{Unit}\\
            \noalign{\smallskip}
            \hline
            \noalign{\smallskip}
1&\mathrm{Origin}&\mathrm{DCMC}&\\
2&\mathrm{Sequence}&\mathrm{JHHMMSS.ss}\pm\mathrm{DDMMSS.s}\\
3,4,5&\mathrm{Ra}&\mathrm{Right\,Ascension\,at\,the\,epoch\,2000}&[h,m,s]\\ 
6,7,8&\mathrm{Dec}&\mathrm{Declination\,at\,the\,epoch\,2000}&[\degr,\arcmin,\arcsec]\\
9&\mathrm{RMS-SYS}&\mathrm{Systematic\,postional\,error}&[\arcsec]\\
10&\mathrm{Xpos-}I&\mathrm{Object\,position\,along\,x}&[pixel]\\
11&\mathrm{Ypos-}I&\mathrm{Object\,position\,along\,y}&[pixel]\\
12&\mathrm{FIELD-POS-}I&\mathrm{Reference\,number\,to\,field\,parameters}&\\
13&\mathrm{STRIP-}I &\mathrm{The\,strip \,number}&\\
14&\mathrm{MAG-APER-}I&\mathrm{Magnitude\,derived\,within\,}7 \arcsec & [mag]\\
15&\mathrm{MAGERR-APER-}I&\mathrm{Magnitude\,error\,derived\,from\,photon\,statistics\,within\,}7 \arcsec &[mag]  \\
16&\mathrm{EXTR-Flag-}I&\mathrm{Extraction\,Flags}&\\
17&\mathrm{IMA-Flag-}I&\mathrm{FLAG-image\,flags\,OR'ed\,over\,the\,iso\,profile}&\\
18&\mathrm{Xpos-}J&\mathrm{Object\,position\,along\,x}&[pixel]\\
19&\mathrm{Ypos-}J&\mathrm{Object\,position\,along\,y}&[pixel]\\
20&\mathrm{FIELD-POS-}J&\mathrm{Reference\,number\,to\,field\,parameters}&\\
21&\mathrm{STRIP-}J&\mathrm{The\,strip \,number}&\\
22&\mathrm{MAG-APER-}J&\mathrm{Magnitude\,derived\,within\,}7\arcsec &[mag]\\
23&\mathrm{MAGERR-APER-}J&\mathrm{Magnitude\,error\,derived\,from\,photon\,statistics\,within\,}7 \arcsec &[mag]\\
24&\mathrm{EXTR-Flag-}J&\mathrm{Extraction\,Flags}&\\
25&\mathrm{IMA-Flag-}J&\mathrm{FLAG-image\,flags\,OR'ed\,over\,the\,iso\,profile}&\\
26&\mathrm{Xpos-}K&\mathrm{Object\,position\,along\,x}&[pixel]\\
27& \mathrm{Ypos-}K&\mathrm{Object\,position\,along\,y}&[pixel]\\
28&\mathrm{FIELD-POS-}K&\mathrm{Reference\,number\,to\,field\,parameters}&\\
29&\mathrm{STRIP-}K&\mathrm{The\,strip \,number}&\\
30&\mathrm{MAG-APER-}K&\mathrm{Magnitude\,derived\,within\,}7 \arcsec &[mag]\\
31&\mathrm{MAGERR-APER-}K&\mathrm{Magnitude\,error\,derived\,from\,photon\,statistics\,within\,}7 \arcsec &[mag]\\
32&\mathrm{EXTR-Flag-}K&\mathrm{Extraction\,Flags}&\\
33&\mathrm{IMA-Flag-}K&\mathrm{FLAG-image\,flags\,OR'ed\,over\,the\,iso\,profile}&\\
34&B\mathrm{PMM}&\mathrm{Blue\,magnitude\,from\,the\,PMM\,catalogue}&\\
35&R\mathrm{PMM}&\mathrm{Red\,magnitude\,from\,the\,PMM\,catalogue}&\\
            \noalign{\smallskip}
            \hline
         \end{array}
      \]
   \end{table*}

  \begin{table}
      \caption[]{Quality table}
         \label{quatab}

      \[
         \begin{array}{clll}
            \hline
            \noalign{\smallskip}
	    \mathrm{Col.}&\mathrm{Name}&\mathrm{Unit}\\
            \noalign{\smallskip}
            \hline
            \noalign{\smallskip}
	    1 &\mathrm{Strip\, Number}   & \\
	    2 &\mathrm{Observing\,date}  & \\
	    3 & I_0                      & \mathrm{[mag]}\\
	    4 & \delta _{I_0}            & \mathrm{[mag]}\\
	    5 & J_0			 & \mathrm{[mag]}\\
	    6 & \delta _{J_0}            & \mathrm{[mag]}\\
            7 & K_0                      & \mathrm{[mag]}\\
            8 & \delta _{K_0}            & \mathrm{[mag]}\\
            9 &\mathrm{Notes}      & \\
             \noalign{\smallskip}
            \hline
         \end{array}
      \]
   \end{table}

\subsubsection{Access}
The DENIS Point Source Catalogue towards the Magellanic Clouds,
data (table \ref{soutab}) and quality (table \ref{quatab}) tables,
are electronically  available  from   CDS  via 
http://\-cds\-web.\-u-\-stra\-sbg.\-fr/\-de\-nis.html.

\section{Concluding remarks}
The catalogue is a suitable tool  for the study of late--type stars in
the  Magellanic   Clouds.   These  studies may  involve  the  statistical
separation of various species of stars, i.e. RGB and AGB (both O--rich
and  C--rich); the  characterization of  the mass  loss  properties of
these stars, when  combined with measurements in the  mid and far--IR;
the relations of infrared colours and magnitudes with variability, when
combined with measurements of light curves (EROS, MACHO) or comparable
photometric    data    (2MASS);     the    interpretation    of    the
Hertzsprung--Russel  diagram through theoretical  evolutionary models;
the investigation of metallicity  effects inside the Magellanic Clouds
and in  comparison with our  own Galaxy; the  study of the  history of
star formation.

\begin{acknowledgements}
The authors kindly thank Ian Glass  for his useful comments and J.L. 
Chevassut, F. Tanguy and K. Weestra for their technical support, 
the  whole DENIS staff and
all the DENIS observers who collected the data.
The DENIS  project is supported by  the SCIENCE and  the Human Capital
and Mobility  plans of the  European Commission under  grants CT920791
and  CT940627  in  France,  by  l'Institut National  des  Sciences  de
l'Univers,  the Minist\`ere  de l'Education  Nationale and  the Centre
National de la  Recherche Scientifique (CNRS) in France,  by the State
of Baden--W\"urttemberg  in Germany,  by the DGICYT  in Spain,  by the
Sterrewacht  Leiden  in  Holland,  by the  Consiglio  Nazionale  delle
Ricerche   (CNR)  in  Italy,   by  the   Fonds  zur   F\"orderung  der
wissenschaftlichen Forschung  and Bundesministerium f\"ur Wissenschaft
und  Forschung in  Austria, by  the Fundation  for the  development of
Scientific Research  of the State  S\~ao Paulo (FAPESP) in  Brazil, by
the OKTA grants F--4239 and F--013990  in Hungary, and by the ESO C \&
EE grant A--04--046.
\end{acknowledgements}

\end{document}